\documentclass[aps,twocolumn,groupedaddress,prb]{revtex4}

\usepackage{graphicx}

\begin{document}

\author{Eric Mandell}
\author{Nathaniel Hunton}
\author{P. Fraundorf}
\email{pfraundorf@umsl.edu}
\affiliation{Department of Physics and Astronomy and Center for Molecular 
Electronics, University of Missouri - StL, St. Louis, Missouri 63121}

\title{Unlayered graphenes in red-giant starsmoke}


\begin{abstract}
Electron diffraction, imaging, and energy 
loss provide evidence for unlayered graphene sheets
in the core of certain interstellar graphite onions 
(from the meteorite Murchison) whose isotopes
indicate formation in the atmosphere of late-stage 
asymptotic giant branch stars (like those 
which nucleo-synthesized much of the earth's carbon).  
The data are compared to structural models loosely associated 
with atom-by-atom, molecule-by-molecule, and dendritic-droplet 
solidification processes.  In this context the observed 
density, diffraction peak-shapes, and edge-on sheet patterns, 
along with theoretical limits on time for growth in the 
presence of outgoing radiation pressure, suggest nucleation 
of hexagonal sheets from pentagons, perhaps from a supercooled 
melt.  These results warrant a closer examination of specimen 
structure, the energetics of unlayered graphene nucleation, 
and processes such as jets in late star atmospheres.
\end{abstract}

\maketitle


\section{Introduction}
\label{sec:Intro}

This paper elaborates on a connection between two forefront 
areas of scientific inquiry:  (i) single-walled carbon 
nanostructure formation, and (ii) the laboratory study of 
astrophysical dust formed around stars other than our own.
In the former context, our observations may be relevant 
specifically to recent work on nucleation processes for 
single-walled carbon nanotubes \citep{Fan03}, as well as 
evidence for the formation of single-walled 
conical \citep{Kasuya02} and multi-walled tubular \citep{deHeer05}
carbon structures from liquid carbon droplets.  

In the latter context, dust particles initially formed in the 
neighborhood of other nucleosynthesis sources, incorporated 
without melting during the formation of our solar system 
into future carbonaceous meteorites, and finally extracted 
for study in laboratories on earth, are proving to be a 
useful source of information on various astrophysical objects 
and processes \citep{Zinner98}.  Along with diamond, silicon carbide, 
and spinel, condensed phases identified as presolar include 
several graphite-related particle types.  In particular, 
micron-sized ``high-density'' and isotopically heavy 
(i.e. with $^{12}$C/$^{13}$C $<$ solar = 89) graphite spheres 
from the Murchison meteorite \citep{Amari95} have ``carbon-black'' 
or ``graphite onion'' {\em rims} characterized by the concentric 
arrangement of 0.34 nm graphite layers.  A sizable fraction of 
these rims appear to have nucleated not on the inorganic 
grains that have been used to constrain condensation 
conditions \citep{Bernatowicz96, Croat03}, but on a spherical 
{\em core} comprised of a novel carbon 
phase \citep{Bernatowicz96, Fraundorf02}.  In addition to its 
relevance to our knowledge of carbon processing in general, 
this phase may provide insight into precipitation processes in 
the atmospheres of AGB stars with surface temperatures low compared to 
our sun \citep{Winter00}, into astrophysical observations of cool 
stellar outflows \citep{Frenklach89,Cherchneff92,Krueger96}, and 
into the state of interstellar condensed 
carbon \citep{Sedlmayr94, Jura97, Chhowalla03}.  

The carbonaceous core phase in these presolar onions is comprised 
of atom-thick ``flakes'' around 4 nm across, with graphite (hk0) 
ordering but with no sign of the 0.34 nm (002) graphite layering 
characteristic of graphite, amorphous carbon, and multiwall carbon 
nanotubes, i.e. of most solid non-diamond carbon 
phases \citep{Bernatowicz95,Bernatowicz96}.  Intersecting line 
pairs in high resolution TEM images suggest that the sheets 
themselves may be bent by the occurence of single cyclopentane 
defects occuring in place of perhaps one of every 400 
graphene hexagons \citep{Fraundorf02}, as though a 
significant fraction of the carbon atoms are in
wide cone-angle faceted carbon nanocones.  How one might pack 
a high abundance of faceted nanocones into a dense solid, while 
supressing graphitic layering, is a mystery for earth-based 
processing.  The fact that it happens in red-giant atmospheres, 
most likely to a significant fraction of the carbon ejected 
into the interstellar medium, only deepens the relevance of 
the mystery.  
 
At one extreme, the cores might form by addition of one carbon 
atom at a time, with layer defects added in randomly from time 
to time to lessen entropy loss during the condensation 
process \citep{Sedlmayr97, Kroto88}.   Alternatively, they might form 
by the ``collisional agglomeration'' of previously-formed 
planar PAHs \citep{Allamandola89, Bernatowicz97c}. 
At the other extreme, the cores might form by the dendritic crystallization 
of liquid carbon rain-drops condensed around magnetohydrodynamic jets
in carbon-rich stellar atmospheres, before ejection 
by radiation pressure into the interstellar medium.  In any of these 
cases, nucleation of the unlayered sheets might 
occur on lower-coordination (e.g. pentagonal) defects, or conversely 
the growing sheets might be converted from two into three dimensional 
structures by defects randomly-generated in the growing sheets.
Observations which bear on these questions are reported in 
separate papers.  Here, we look at what structural models linked
loosely to these processes have to say in context of such 
observational data.

\section{Models}
\label{sec:Models}

\begin{figure}[tbp]
\includegraphics[scale=.7]{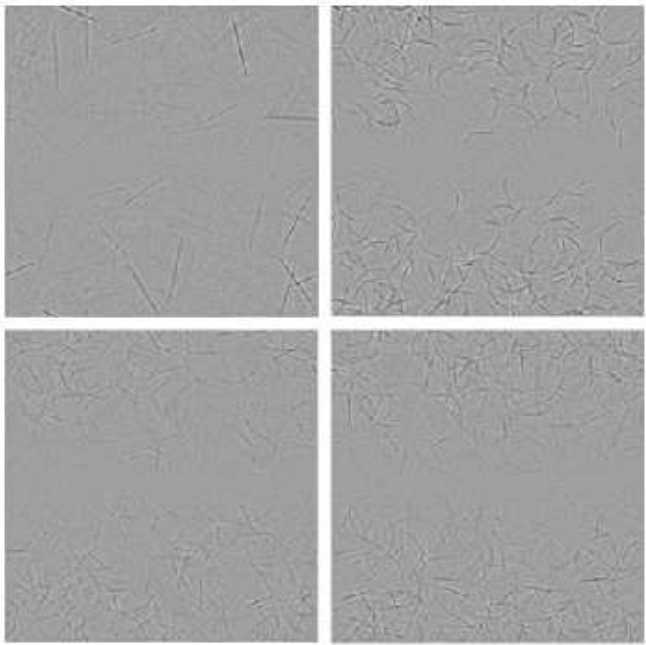}%
\caption{Randomly-positioned graphene sheets (upper left),
relaxed pentacones (upper right), faceted pentacones 
(lower left) and pentacones with two adjacent sheets 
parallel (lower right), seen in HRTEM image simulations 
of a specimen 22.5 nm square and 10 nm thick with a 
horizontal perforation.}
\label{Fig90}
\end{figure}

Other papers associated with this work 
will focus first on the experiments.  Here 
we begin with a set of simple models, loosely 
connected to the formation processes mentioned 
above.  The models will be illustrated (Fig. \ref{Fig90}) 
by high resolution electron microscope images, 
to which they'll eventually be compared.  

\subsection{Agglomerated PAH's}
\label{subsec:Sheets}

One process suggested for the formation of these 
unlayered graphene cores has been the collisional aggregation 
of polycyclic aromatic hydrocarbon (PAH) molecules.  They've 
been detected in astrophysical settings, and (as we 
confirm) they could explain the major features seen in 
electron diffraction patterns.  Our first model for this 
is randomly-oriented graphene sheets like the 640 atom
sheets illustrated in the upper left of Fig. \ref{Fig90}.

The astrophysical difficulty associated with this, 
as a source of the graphene cores, is that incoming molecules 
are likely to have significant velocities (due to 
thermal, orbital, and radiation-pressure processes) 
relative to any prospective agglomerate.  
Even formation of large porous clusters seems difficult,
given the expected kinetic energy of new arrivals.  
Formation of spherical particles whose density 
is comparable to that of graphite seems even less likely.  
Nonetheless, comparing the core material with randomly-oriented 
graphene sheets should be enlightening.

\subsection{Corranulene Spirals}
\label{subsec:WeberCones}

A second alternative is growth of unlayered graphenes 
from the vapor phase.  PAH's form in space so this is 
certainly possible, provided one can prod the particles 
to grow in three dimensions rather than as sheets, without
resorting to the van der Waals layering process 
(graphetization) so ubiquitous in non-diamond carbon on 
earth.  The possibility of doing this with help from an 
occasional pentagon or heptagon in the sheet has already
been proposed \citep{Kroto88}.  Although we are looking 
into atom-by-atom growth models as well, we simulate 
this alternative here with the random collection of 
relaxed (circular cross-section) pentacones shown 
in the upper right of Figure \ref{Fig90}.  These 
pentacones are relaxed because we expect that in vacuo 
the energy preference for sheet flatness will 
force all carbon atoms to share some of the curvature, 
at least as sheet coherence widths approach the 4 nm
values seen in these specimens.

If growth occurs atom-by-atom, it is likely nucleation 
limited and the center of the particle might show some 
special features (not yet reported).  There is also the 
mystery about suppressed graphite layering.  If we 
have difficulty suppressing this layering even in amorphous 
carbon on earth, can radiation or some other process 
in the star's atmosphere allow graphene sheets to grow 
edge-wise without allowing atoms to start new 
van der Waals layered sheets as well.  Although the 
energetic difference between such sites makes this possible, 
no quantitative process (experimental or theoretical) has 
been demonstrated to suppress layering so far.

\subsection{Dendritic Solidification}
\label{subsec:FacetedCones}

A third alternative is dendritic solidification of a 
liquid carbon drop.  Advantages of this are that it 
explains observed densities \citep{Ghiringhelli05} and the 
spherical core shape.  There is an astrophysical 
problem with the sticking coefficient of incoming 
atoms \citep{Michael03} that may also be ameliorated if the target 
cluster is liquid.  As mentioned above, even though 
liquid carbon cannot equilibrate at low pressure 
it has been implicated in the laboratory formation 
of both single walled nanohorns and carbon nanotubes 
in low-pressure non-equilibrium processes.  The high 
energy of sheet-edge attachment and proximity of 
atoms in the liquid might easily explain why graphene 
layering is suppressed as well.  We also expect 
growth of sheets in a melt (possibly from pentagonal 
nuclei) would result in faceted nanocones since interstitial 
carbon would leave no room to relax.  Thus our model 
for this is simply the faceted pentacones shown in 
the lower part of Figure \ref{Fig90}.  On the lower left, 
you'll find symmetric faceted cones like those you might 
expect if the pentagon is the nucleus.  If pentagons
also arise spontaneously during growth, flat growth 
might be redirected into three (not four) additional 
directions resulting in the asymmetric nanocones 
shown on the lower right.  

The disadvantage of this model is glaring.  No one 
expects it.  Since we have no direct experimental 
access to stellar atmospheres cool enough to condense 
carbon, this may be a case of finding the raindrop 
before we hear the storm, but its unexpectedness 
also means that more evidence for such processes will be 
needed to make it convincing.  That's why the 
focus of this paper has been on listening to what the
specimens have to say.

\section{Experimental}
\label{sec:Exp}

\begin{figure}[tbp]
\includegraphics[scale=1]{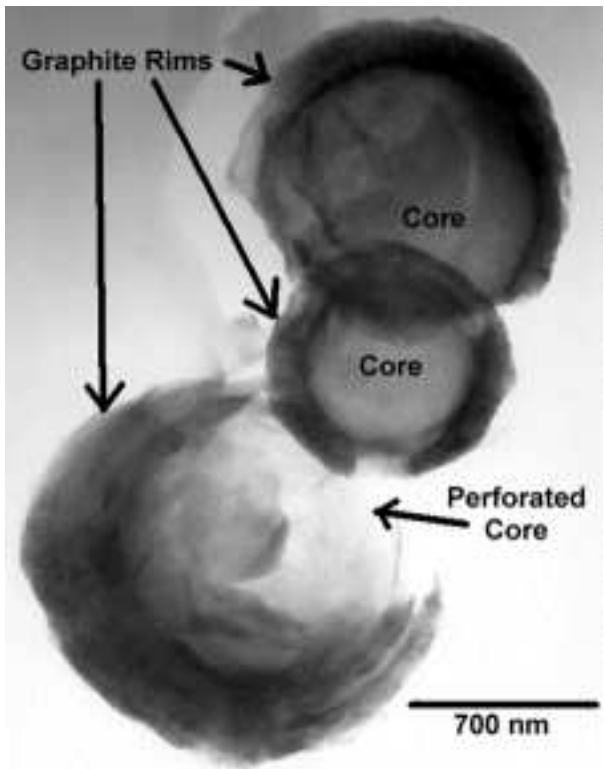}%
\caption{Image of three onion slices, showing one with a torn core suitable
for electron phase contrast study.}
\label{Fig1}
\end{figure}

Figure \ref{Fig1} is an image of several 
graphite onions with an unlayered graphene core.  They 
were from the Murchison meteorite graphite separate KFC1 (2.15-2.20 g/cm$^3$), whose preparation has been described in detail by \citet{Amari94}. The onions were microtomed to a thickness of 70 nm, and deposited onto 3mm copper grids with a holey carbon support film, by \citet{Bernatowicz96}. Specimens 
were examined in a 300 kV Philips EM430ST TEM with point 
resolution near $0.2 nm$.  

Cleanly-sectioned cores large 
enough to minimize ``spillover diffraction'' from surrounding rims were the 
primary target for selected area diffraction measurements (c.f. Fig \ref{Fig1}) 
on a TEM specimen designated KFC1A:E.  Electron energy loss density 
measurements, and energy filtered maps, were taken on similar 
specimens.  On the other hand, electron phase contrast (HRTEM) 
images focused instead on onions whose cores were torn during the 
sectioning, and positioned over a hole in the supporting carbon film.  
Image simulations were done using the strong phase object approximation.  
Further details of the data acquisition strategy are discussed 
elsewhere.

\section{Diffraction Data}
\label{sec:Diffraction}

\begin{figure}[tbp]
\includegraphics[scale=0.75]{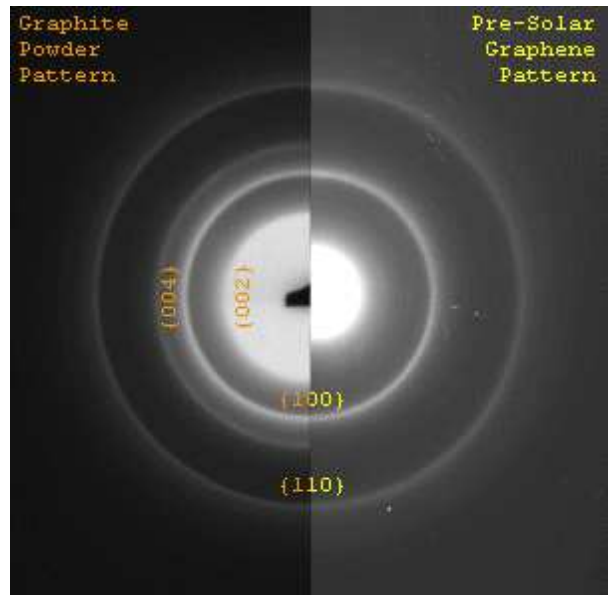}%
\caption{Experimental diffraction pattern of polycrystalline 
graphite (left) and presolar unlayered graphene (right).  Note 
the absence of all except (hk0) spacings in the latter, as 
well as the high frequency tails on each ring.}
\label{Fig2b}
\end{figure}

\begin{figure}[tbp]
\includegraphics[scale=0.75]{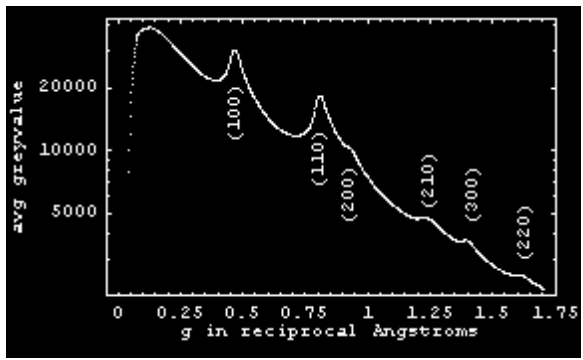}%
\caption{Log plot of azimuthally-averaged 
greyvalue from a graphite-onion core, 
showing the presence only of ``in-sheet'' graphite 
(hk0) spacings out to 1.7 reciprocal Angstroms.}
\label{Fig2c}
\end{figure}

Figure \ref{Fig2b} compares the diffraction pattern 
of an onion core (right half) to that from 
graphetized carbon.  The (hk0) peaks in both have 
``high frequency halos'', in the latter case because
the carbon layers have graphitic onion morphology 
which makes it impossible for adjacent graphene 
sheets to retain coherence from layer to layer.  
Figure \ref{Fig2c} shows one aziumuthally-averaged 
``electron powder diffraction profile'' from an 
onion core on a log-intensity scale.  As you can 
see, all graphite reflections except for the six expected graphene (hk0) 
spacings are systematically absent from the 
patterns.  This is true even for the graphite 
(002) spacing near 0.29[$\AA^{-1}$], easily detected 
in both diffraction and HRTEM imaging, and present in 
much terrestrial non-diamond carbon (including evaporated 
``amorphous''carbon support films,
single-walled nanohorn \citep{Bandow00} and 
nanotube collections because 
of the propensity of 
such tubes to align with van der Waals bonding between 
adjacent tube walls). 

\subsection{Flat sheet profiles}
\label{subsec: Warren}

\begin{figure}[tbp]
\includegraphics[scale=0.7]{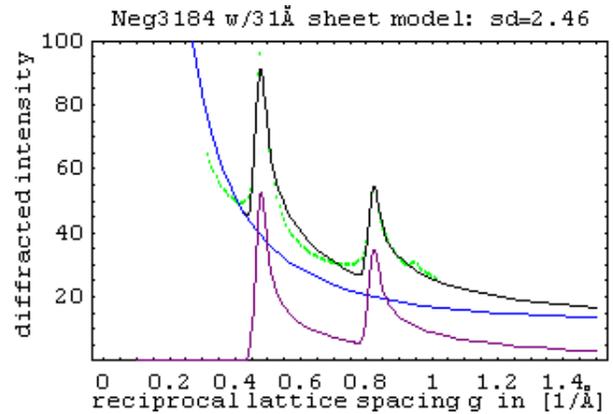}%
\caption{Least-squares fit of core pattern 3184 
(green) with a single-size sheet model (black) 
has the smallest standard deviation (2.46 greyvalue 
units) using 3.1[nm] sheets.}
\label{Fig3}
\end{figure}

Azimuthally-averaged peak profiles like that in 
Fig. \ref{Fig3} allow one to see the asymmetric 
peak broadening (i.e. halos on the high-frequency 
side) characteristic of atom-thick 
crystals that continue to diffract even when 
highly foreshortened by tilt with respect to the 
incident beam.  As a result, the product of 
peak full-width at half-maximum, and grain 
size is, is closer to 2 ($\cong 1.84$) \citep{Warren41} 
than to values near 1 typical of equant 3D crystal 
shapes \citep{Patterson39b}.  This rule estimates 
mass-weighted average sheet sizes of 3.8 to 4[nm] 
for the (110) and (100) peaks of Figure \ref{Fig3}, 
in agreement with sizes inferred from other onion 
core patterns by \citet{Bernatowicz96}.  Further, 
comparable widths for the (110) and (100) reflections 
argues against systematic in-sheet anisotropy \citep{Rees50}.

The Figure \ref{Fig3} analysis was done by 
adding random layer lattice profiles from Warren \citep{Warren41} 
for (100) and (110) rows of a single-size graphene 
sheet, plus a simple $a + b/g^2$ background, to 
azimuthally-averaged diffraction profiles whose 
intensity varied linearly with electron exposure.  
The consistency in peak asymmetry 
between experiment and theory (i.e. the common 
high-frequency tail), and the 
absence of graphite layering spacings like (002) in our experimental 
patterns, support the interpretation by \citet{Bernatowicz96} 
that diffraction from this core material is indeed dominated by 
randomly-oriented atom-thick graphene sheets.  
The fit in Figure \ref{Fig3} is also shows that a 
single sheet-size alone does not explain details at 
the base of these peaks.  

Similar fits were attempted using two different 
sheet sizes, or a log-normal distribution, with 
limited success.  In other 
words, problems at the base of each peak (like 
the sharp cusp on the right side of the valley between 
both large peaks) remained.  
At this point, it began to 
seem that the feature responsible for broadening
the base of the peak does not carry with it 
the asymmetry expected for sheets 
of atom-thick graphene, even though feature 
locations indicate graphene-like spacings.

\begin{figure}[tbp]
\includegraphics[scale=0.7]{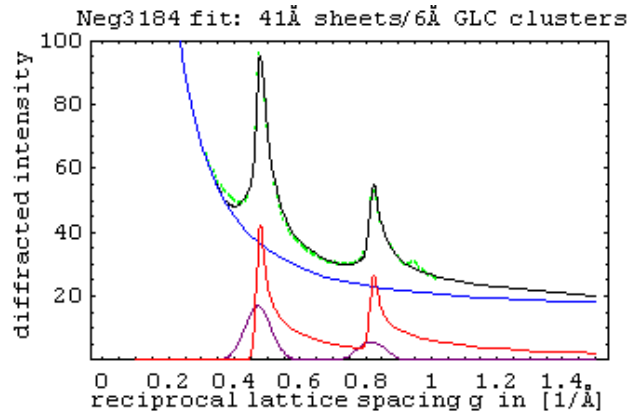}%
\caption{Least-squares fit of core pattern 3184 
(green) with a two-component mix (black) of graphene sheets and 
graphene-like carbon clusters has the smallest 
standard deviation (0.98 greyvalue units) for 
4.1[nm] sheets and 0.6[nm] clusters.}
\label{Fig5}
\end{figure}

Figure \ref{Fig5} fits a two-parameter 
model that combines graphene sheets of 
one size with broader gaussian peaks at 
the graphene peak locations to one of 
several experimental spectra analyzed in 
this way.  In each case the the systematic 
peak-shape problem was noticeably 
reduced, and the fit quality 
improved as well. 

Instrumental broadening of peaks cannot 
explain the symmetric base-broadening 
component for two reasons.  First, patterns 
from polycrystalline Al under the same 
illumination conditions limit the 
instrumental broadening full-width at 
half-maximum to the 0.01 reciprocal 
Angstrom range, much less than the 
0.16 reciprocal Angstrom width of the 
symmetric residual.  Secondly, the 
sharp leading edge of the asymmetric 
peaks of each core profile itself limits 
instrumental broadening to well under 
0.05 reciprocal Angstroms.

Such peaks might also be due to ``graphene-like 
carbon'' clusters whose peaks do 
not show the asymmetry of atom-thick sheets.  
Since diamond also has essentially the 
same C-C bond lengths and hence first 
two reflections, one might also think of 
these as ``diamond-like carbon'' peaks 
although the lack of higher-order 
diamond reflections in Fig. \ref{Fig2c} 
like those from pre-solar nanodiamonds \citep{Fraundorf89},  
and plasmon energies seen in electron energy loss 
spectra \citep{Bernatowicz96},
limit the extent to which carbon-$sp^3$ ordering is 
actually present.

\subsection{Correlated Sheet Profiles}
\label{subsec: debye}

\begin{figure}[tbp]
\includegraphics[scale=0.7]{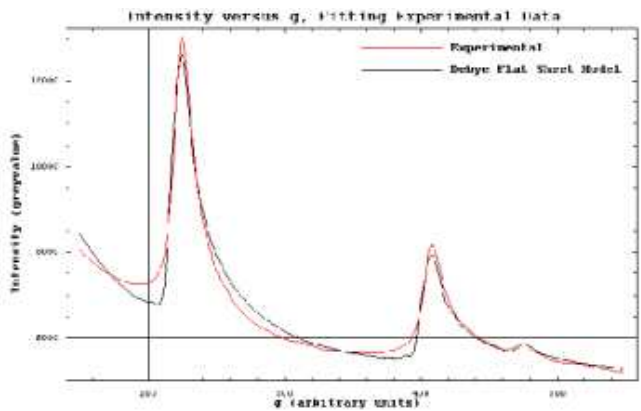}%
\caption{Experimental diffraction data from a presolar core plotted 
with the best fit provided by a flat sheet Debye model.}
\label{ericdebye}
\end{figure}

To consider more complicated effects, profiles for 
unlayered graphene molecules were also
calculated using the Debye scattering equation \citep{Warren69}.  
Here, the diffraction intensity at each spatial frequency is 
computed by summing over all atoms in the structure.  Since the Debye 
model works for any list of atom positions (given enough computer time), 
powder diffraction profiles for any list of atom positions may be computed from 
a sum over atom pairs.  By creating a 
sheet-size fitting routine, we let the computer choose the best single-sheet size fit to 
experimental data.  Figure \ref{ericdebye} illustrates that the flat-sheet results are the 
same as those seen in the Warren model, where scattering on the low-frequency side 
and at the graphene periodicities remains unexplained.

\begin{figure}[tbp]
\includegraphics[scale=0.7]{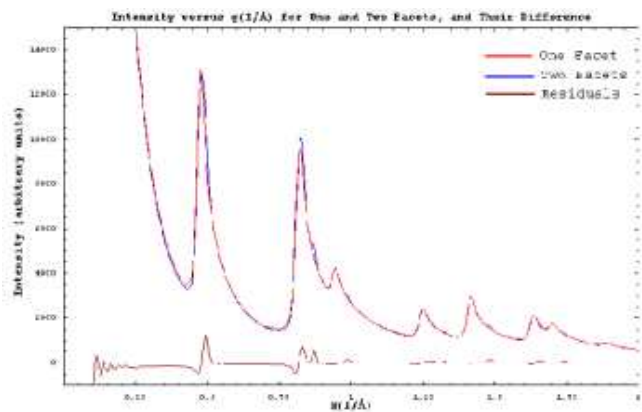}%
\caption{A comparison of Debye diffraction patterns for one and two facets of the faceted nanocone.  Observe the occurrence of a satellite peak on the (110) high frequency tail, due to the orientation of the facets.}
\label{ericdebye2}
\end{figure}

A possible explanation for differences between the flat sheet model and experimental data are an effect which arises from the wide-range 
fringe visibilty of atom-thick crystals \citep{Fraundorf05JAP}, 
namely coherence effects from coordination between sheets.  
The Debye model (unlike Warren) lets us examine the effects of 
such sheet-sheet correlations.
This can be seen in profile by comparing Debye models for one sheet of a faceted carbon nanocone with that for two adjoined facets, which make some non-zero angle with one another, where the profiles have been scaled for scattering by the same number of atoms.  Figure \ref{ericdebye2} shows the profiles and their difference, where there is more scattering at the (110) peak due to the increased coherence width of that spacing across both facets, and a small satellite peak, corresponding to the preferential projection of this spacing across the facets, due to the angle between the two sheets.  If we were to take many facets at different angles, the satellite peaks would blur, yet we would see a small amount of "extra" scattering at the graphene periodicities.  Early evidence from Debye work on collections of randomly-oriented multi-sheet structures (e.g. pentacones) suggests that coherence effects provide features similar to 
the differences seen between experimental data and single sheet models.

\section{Density Measurements}
\label{sec:Density}

Using a Leo-Zeiss 912 Energy Filtered TEM, we have obtained both brightfield TEM images, and zero-loss images of presolar graphite spherules.  The brightfield images are typical unfiltered, phase contrast TEM images, while the zero-loss images are formed only with electrons which have undergone elastic scattering.   The inelastic mean-free path is defined as the e-folding distance for zero-loss intensity with specimen thickness, and this in turn is proportional to projected potential and (for regions of similar composition) to mass densities in grams per cm$^2$.  Thus the ratio between inelastic mean free path at two locations (e.g. between the rim and core of a constant thickness slice) will to first order give us a ratio of mass densities per unit volume ($\rho$).

\begin{widetext}
\begin{figure}[tbp]
\includegraphics[scale=0.95]{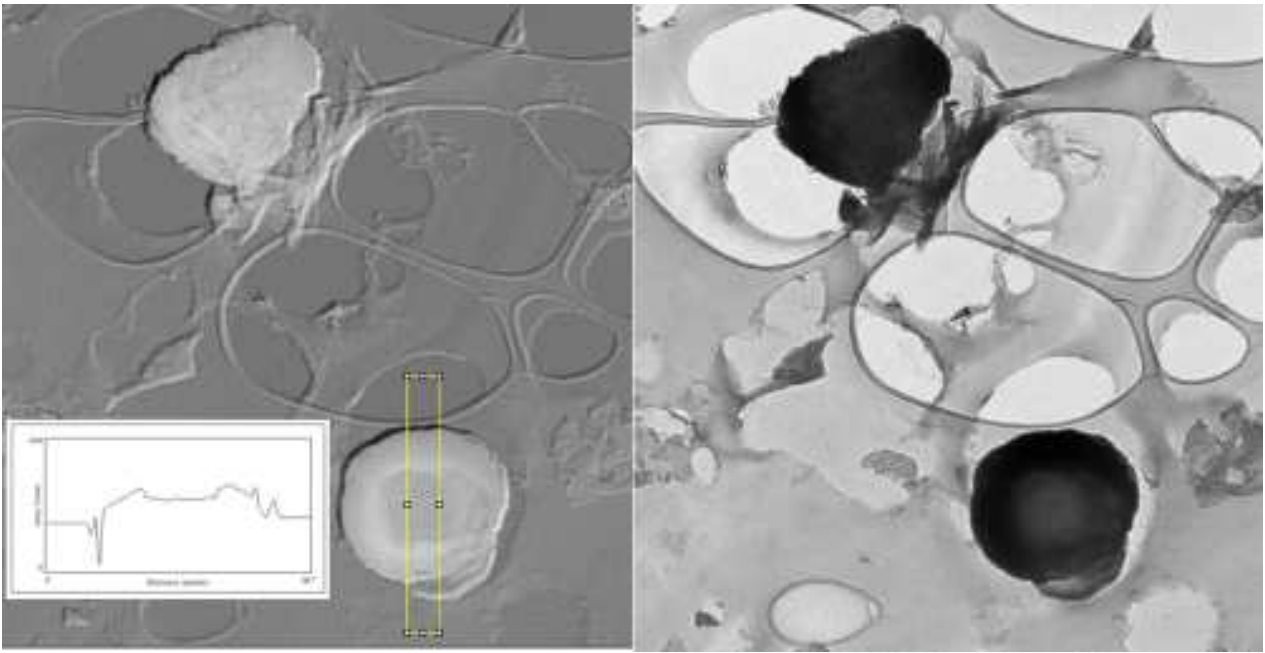}%
\caption{A mean-free-path "thickness" map (a) calculated from a brightfield TEM image (b).  The resulting greyvalue intensity profile (c) is shown for the windowed section in the "thickness" map.  Intensities for the rim and core, relative to the hole, can be used to gauge density changes between the two regions.}
\label{ericeftem}
\end{figure}
\end{widetext}

Serial EELS spectra on separate core and rim regions, obtained with a Gatan 607 series spectrometer, has suggested core densities around 0.65 relative to the graphitic rim ($\rho \simeq 2.2$ g/cm$^3$).  However, the microtome does not always slice uniformly across an onion.  Therefore we also obtained "thickness" images using the log ratios of brightfield and elastic images taken with an energy-filtered TEM.  Images were acquired with a CCD camera with no changes in illumination conditions between the two images, other than the energy filtering for the elastic image.  Averaged intensity profiles are then taken over windowed sections of the resultant "thickness" map.  Greyvalue intensities for the core and rim can be compared, relative to the greyvalue intensity of a hole in the image.  Figure \ref{ericeftem} shows the thickness map for two graphite spherules and the regular brightfield image for comparison.  We concentrate on the spherule showing a strong core-rim structure, and calculate a greyvalue profile over the selected area shown.  Comparing intensities for the rim and core, relative to the hole, intensity, we find core to rim thickness ratios of 0.64.  Thus the core material seems to have a density near 1.5 grams per cm$^3$, i.e. lower than the 1.69 grams per cm$^3$ of buckminsterfullerene.  This should be considered when comparing between the various proposed growth models, and further emphasizes the difference between presolar onion core and rim.

\section{High Resolution Imaging}
\label{sec:Imaging}

HRTEM work on core material \citep{Fraundorf02} has 
suggested the existence of coordinated graphene sheets, seen as 
end-connected linear features in images.  Simulations (cf. Figure \ref{Fig90}) 
show that edge-on graphene sheets, often intersecting at a 
point, indeed dominate the contrast.  These simulations also 
show that relaxed nanocones (upper right) give noticably stronger contrast 
fluctuations than do faceted cones or single sheets with the
same number of atoms.  Relaxed nanocones also give an 
appearance of curvature not reported in experimental images \citep{Fraundorf02}, 
although more quantitative work is still needed to see if the experimental 
images indicate nanocone faceting, and perhaps even provide 
evidence for nucleation on pentagons.  

\begin{figure}[tbp]
\includegraphics[scale=.7]{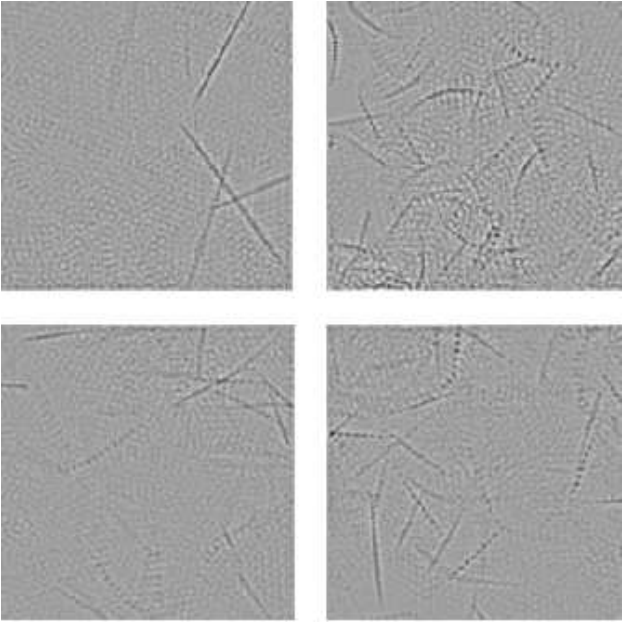}%
\caption{Randomly-positioned graphene sheets (upper left),
relaxed pentacones (upper right), faceted pentacones 
(lower left) and pentacones with two adjacent sheets 
parallel (lower right), seen in HRTEM image simulations 
at 1 $\AA$ resolution.}
\label{Fig91}
\end{figure}

Future work will be assisted by the availability of aberration-corrected
microscopes.  Figure \ref{Fig91} shows the center region of Figure \ref{Fig90} 
as it would look in a microscope with twice the point resolution and 
a correspondingly better damping envelope.  The biggest gain may come 
from an ability to see atom columns in the edge-on sheets, so that the 
coherence of intersecting segments can be investigated.  Contrast 
fluctuations are also enhanced at higher resolution, so quantitative analysis 
of fluctuations should reveal new information as well.

\section{Discussion}
\label{sec:Discuss}


The data here confirms previous observations 
but shows that randomly-oriented flat sheets 
are not adequate to explain the diffraction data.  
It also indicates that the density of the 
core material is less than 2 grams per cm$^3$, 
closer to that expected for 
frozen liquid carbon than for graphite, 
and shows how high resolution TEM imaging 
should be able to distinguish between 
single sheets, faceted pentacones, and 
relaxed pentacones.  It also connects these 
structural alternatives to three different 
types of formation process:  sheet agglomeration, 
atom-by-atom growth, or dendritic solidification 
of a liquid drop.  

Other mysteries include the abrupt transition from unlayered 
graphene to normal onion growth (Fig. \ref{Fig1}), 
the low ``average'' density of gas 
in grain-forming regions of red giant 
atmospheres which predicts too little time for 
growth prior to ejection by radiation pressure \citep{Bernatowicz96}, 
and the low sticking coefficient for carbon 
atoms incident onto solid surfaces from the 
gas phase \citep{Michael03}.  We are separately working to confirm 
interpretation of existing HRTEM data that 
suggests that ``flat'' graphene sheets \citep{Liu94}
connect pentagonal sheet defects, instead 
of the curved sheets one might expect if  
free-standing single-walled nanocones were allowed 
to relax before incorporation into the solid. 

Might undercooled liquid-carbon droplets \citep{Kasuya02, deHeer05}  
crystallize by dendritic growth of 
individual graphene sheets?   The proximity of 
liquid atoms on both sides of a growing 
sheet could prevent both van der Waals 
layering, and sheet relaxation to accomodate 
strains caused by the occasional sheet 
defects which perturb sheet growth into new 
directions.  Nucleation of 
this melt directly on pentagons might be 
consistent with recent studies of carbon 
nanotube energetics \citep{Fan03}.  Liquid 
carbon atoms left between ``superstructure'' 
sheets might have little choice but to 
form unsheetlike ``gap-filling'' structures 
on final solidification.  The 
spherical core shape and density, 
a high sticking coefficient for incoming 
carbon atoms, and the abrupt transition to 
graphitic onion growth after solidification 
could all be explained nicely as well.  
Of course, if this unexpected conclusion 
pans out, the study of weather in cool 
stellar atmospheres may have to open up 
a new chapter on carbon rain.

\begin{acknowledgments}
Thanks to Tom Bernatowicz, Kevin Croat, and 
Roy Lewis for trusting us with specimens of unsliced and 
sliced graphite onions from Murchison for this 
work, and to Howard Berg for access to that 
energy-filtered TEM.  This work has also benefited indirectly 
from support by the U.S. Department of Energy, 
the Missouri Research Board, as well as 
Monsanto and MEMC Electronic Materials Companies. 
\end{acknowledgments}

\bibliography{etmrefs}

\begin{thebibliography}{31}
\expandafter\ifx\csname natexlab\endcsname\relax\def\natexlab#1{#1}\fi
\expandafter\ifx\csname bibnamefont\endcsname\relax
  \def\bibnamefont#1{#1}\fi
\expandafter\ifx\csname bibfnamefont\endcsname\relax
  \def\bibfnamefont#1{#1}\fi
\expandafter\ifx\csname citenamefont\endcsname\relax
  \def\citenamefont#1{#1}\fi
\expandafter\ifx\csname url\endcsname\relax
  \def\url#1{\texttt{#1}}\fi
\expandafter\ifx\csname urlprefix\endcsname\relax\def\urlprefix{URL }\fi
\providecommand{\bibinfo}[2]{#2}
\providecommand{\eprint}[2][]{\url{#2}}

\bibitem[{\citenamefont{Fan et~al.}(2003)\citenamefont{Fan, Buczko, Puretzky,
  Geohegan, Howe, Pantelides, and Pennycook}}]{Fan03}
\bibinfo{author}{\bibfnamefont{X.}~\bibnamefont{Fan}},
  \bibinfo{author}{\bibfnamefont{R.}~\bibnamefont{Buczko}},
  \bibinfo{author}{\bibfnamefont{A.~A.} \bibnamefont{Puretzky}},
  \bibinfo{author}{\bibfnamefont{D.~B.} \bibnamefont{Geohegan}},
  \bibinfo{author}{\bibfnamefont{J.~Y.} \bibnamefont{Howe}},
  \bibinfo{author}{\bibfnamefont{S.~T.} \bibnamefont{Pantelides}},
  \bibnamefont{and} \bibinfo{author}{\bibfnamefont{S.~J.}
  \bibnamefont{Pennycook}}, \bibinfo{journal}{Phys. Rev. Lett.}
  \textbf{\bibinfo{volume}{90}}(\bibinfo{number}{14}), \bibinfo{pages}{145501}
  (\bibinfo{year}{2003}).

\bibitem[{\citenamefont{Kasuya et~al.}(2002)\citenamefont{Kasuya, Yudasaka,
  Takahashi, Kokai, and Iijima}}]{Kasuya02}
\bibinfo{author}{\bibfnamefont{D.}~\bibnamefont{Kasuya}},
  \bibinfo{author}{\bibfnamefont{M.}~\bibnamefont{Yudasaka}},
  \bibinfo{author}{\bibfnamefont{K.}~\bibnamefont{Takahashi}},
  \bibinfo{author}{\bibfnamefont{F.}~\bibnamefont{Kokai}}, \bibnamefont{and}
  \bibinfo{author}{\bibfnamefont{S.}~\bibnamefont{Iijima}},
  \bibinfo{journal}{J. Phys. Chem B} \textbf{\bibinfo{volume}{106}},
  \bibinfo{pages}{4947} (\bibinfo{year}{2002}).

\bibitem[{\citenamefont{de~Heer et~al.}(2005)\citenamefont{de~Heer, Poncharal,
  Berger, Gezo, Song, Bettini, and Ugarte}}]{deHeer05}
\bibinfo{author}{\bibfnamefont{W.~A.} \bibnamefont{de~Heer}},
  \bibinfo{author}{\bibfnamefont{P.}~\bibnamefont{Poncharal}},
  \bibinfo{author}{\bibfnamefont{C.}~\bibnamefont{Berger}},
  \bibinfo{author}{\bibfnamefont{J.}~\bibnamefont{Gezo}},
  \bibinfo{author}{\bibfnamefont{Z.}~\bibnamefont{Song}},
  \bibinfo{author}{\bibfnamefont{J.}~\bibnamefont{Bettini}}, \bibnamefont{and}
  \bibinfo{author}{\bibfnamefont{D.}~\bibnamefont{Ugarte}},
  \bibinfo{journal}{Science} \textbf{\bibinfo{volume}{307}},
  \bibinfo{pages}{907} (\bibinfo{year}{2005}).

\bibitem[{\citenamefont{Zinner}(1998)}]{Zinner98}
\bibinfo{author}{\bibfnamefont{E.}~\bibnamefont{Zinner}},
  \bibinfo{journal}{Ann. Rev. Earth Planet. Sci.}
  \textbf{\bibinfo{volume}{26}}, \bibinfo{pages}{147} (\bibinfo{year}{1998}).

\bibitem[{\citenamefont{Amari et~al.}(1995)\citenamefont{Amari, Zinner, and
  Lewis}}]{Amari95}
\bibinfo{author}{\bibfnamefont{S.}~\bibnamefont{Amari}},
  \bibinfo{author}{\bibfnamefont{E.}~\bibnamefont{Zinner}}, \bibnamefont{and}
  \bibinfo{author}{\bibfnamefont{R.~S.} \bibnamefont{Lewis}},
  \bibinfo{journal}{Meteoritics} \textbf{\bibinfo{volume}{30}},
  \bibinfo{pages}{679} (\bibinfo{year}{1995}).

\bibitem[{\citenamefont{Bernatowicz et~al.}(1996)\citenamefont{Bernatowicz,
  Cowsik, Gibbons, Lodders, Jr., Amari, and Lewis}}]{Bernatowicz96}
\bibinfo{author}{\bibfnamefont{T.}~\bibnamefont{Bernatowicz}},
  \bibinfo{author}{\bibfnamefont{R.}~\bibnamefont{Cowsik}},
  \bibinfo{author}{\bibfnamefont{P.~C.} \bibnamefont{Gibbons}},
  \bibinfo{author}{\bibfnamefont{K.}~\bibnamefont{Lodders}},
  \bibinfo{author}{\bibfnamefont{B.~F.} \bibnamefont{Jr.}},
  \bibinfo{author}{\bibfnamefont{S.}~\bibnamefont{Amari}}, \bibnamefont{and}
  \bibinfo{author}{\bibfnamefont{R.~S.} \bibnamefont{Lewis}},
  \bibinfo{journal}{Astrophysical Journal} \textbf{\bibinfo{volume}{472}},
  \bibinfo{pages}{760} (\bibinfo{year}{1996}).

\bibitem[{\citenamefont{Croat et~al.}(2003)\citenamefont{Croat, Bernatowicz,
  Amari, Messenger, and Stadermann}}]{Croat03}
\bibinfo{author}{\bibfnamefont{T.~K.} \bibnamefont{Croat}},
  \bibinfo{author}{\bibfnamefont{T.}~\bibnamefont{Bernatowicz}},
  \bibinfo{author}{\bibfnamefont{S.}~\bibnamefont{Amari}},
  \bibinfo{author}{\bibfnamefont{S.}~\bibnamefont{Messenger}},
  \bibnamefont{and} \bibinfo{author}{\bibfnamefont{F.~J.}
  \bibnamefont{Stadermann}}, \bibinfo{journal}{Geochim. et. Cosmochim. Acta}
  \textbf{\bibinfo{volume}{67}}, \bibinfo{pages}{4705} (\bibinfo{year}{2003}).

\bibitem[{\citenamefont{Fraundorf and Wackenhut}(2002)}]{Fraundorf02}
\bibinfo{author}{\bibfnamefont{P.}~\bibnamefont{Fraundorf}} \bibnamefont{and}
  \bibinfo{author}{\bibfnamefont{M.}~\bibnamefont{Wackenhut}},
  \bibinfo{journal}{Ap. J. Lett.} \textbf{\bibinfo{volume}{578}},
  \bibinfo{pages}{L153} (\bibinfo{year}{2002}).

\bibitem[{\citenamefont{Winters et~al.}(2000)\citenamefont{Winters, Bertre,
  Jeong, Helling, and Sedlmayr}}]{Winter00}
\bibinfo{author}{\bibfnamefont{J.~M.} \bibnamefont{Winters}},
  \bibinfo{author}{\bibfnamefont{T.~L.} \bibnamefont{Bertre}},
  \bibinfo{author}{\bibfnamefont{K.~S.} \bibnamefont{Jeong}},
  \bibinfo{author}{\bibfnamefont{C.}~\bibnamefont{Helling}}, \bibnamefont{and}
  \bibinfo{author}{\bibfnamefont{E.}~\bibnamefont{Sedlmayr}},
  \bibinfo{journal}{Astron. Astrophys.} \textbf{\bibinfo{volume}{361}},
  \bibinfo{pages}{641} (\bibinfo{year}{2000}).

\bibitem[{\citenamefont{Frenklach and Feigelson}(1989)}]{Frenklach89}
\bibinfo{author}{\bibfnamefont{M.}~\bibnamefont{Frenklach}} \bibnamefont{and}
  \bibinfo{author}{\bibfnamefont{E.~D.} \bibnamefont{Feigelson}},
  \bibinfo{journal}{Ap. J.} \textbf{\bibinfo{volume}{341}},
  \bibinfo{pages}{372} (\bibinfo{year}{1989}).

\bibitem[{\citenamefont{Cherchneff et~al.}(1992)\citenamefont{Cherchneff,
  Barker, and Tielens}}]{Cherchneff92}
\bibinfo{author}{\bibfnamefont{I.}~\bibnamefont{Cherchneff}},
  \bibinfo{author}{\bibfnamefont{J.~R.} \bibnamefont{Barker}},
  \bibnamefont{and} \bibinfo{author}{\bibfnamefont{A.~G. G.~M.}
  \bibnamefont{Tielens}}, \bibinfo{journal}{Ap. J.}
  \textbf{\bibinfo{volume}{401}}, \bibinfo{pages}{269} (\bibinfo{year}{1992}).

\bibitem[{\citenamefont{Krueger et~al.}(1996)\citenamefont{Krueger, Patzer, and
  Sedlmayr}}]{Krueger96}
\bibinfo{author}{\bibfnamefont{D.}~\bibnamefont{Krueger}},
  \bibinfo{author}{\bibfnamefont{A.~B.~C.} \bibnamefont{Patzer}},
  \bibnamefont{and} \bibinfo{author}{\bibfnamefont{E.}~\bibnamefont{Sedlmayr}},
  \bibinfo{journal}{Astron. Astrophys.} \textbf{\bibinfo{volume}{313}},
  \bibinfo{pages}{891} (\bibinfo{year}{1996}).

\bibitem[{\citenamefont{Sedlmayr}(1994)}]{Sedlmayr94}
\bibinfo{author}{\bibfnamefont{E.}~\bibnamefont{Sedlmayr}}, in
  \emph{\bibinfo{booktitle}{Molecules in the Stellar Environment}}, edited by
  \bibinfo{editor}{\bibfnamefont{U.~G.} \bibnamefont{Jorgensen}}
  (\bibinfo{publisher}{Springer-Verlag}, \bibinfo{address}{Berlin},
  \bibinfo{year}{1994}), no. \bibinfo{number}{146} in \bibinfo{series}{IAU
  Colloquium}, pp. \bibinfo{pages}{163--185}.

\bibitem[{\citenamefont{Jura}(1997)}]{Jura97}
\bibinfo{author}{\bibfnamefont{M.}~\bibnamefont{Jura}}, in
  \emph{\bibinfo{booktitle}{Astrophysical implications of the laboratory study
  of presolar materials}}, edited by \bibinfo{editor}{\bibfnamefont{T.~J.}
  \bibnamefont{Bernatowicz}} \bibnamefont{and}
  \bibinfo{editor}{\bibfnamefont{E.}~\bibnamefont{Zinner}}
  (\bibinfo{publisher}{American Institute of Physics},
  \bibinfo{address}{Woodbury NY}, \bibinfo{year}{1997}), no.
  \bibinfo{number}{402} in \bibinfo{series}{AIP Conference Proceedings}, pp.
  \bibinfo{pages}{379--390}.

\bibitem[{\citenamefont{Chhowalla et~al.}(2003)\citenamefont{Chhowalla, Wang,
  Sano, Teo, Lee, and Amaratunga}}]{Chhowalla03}
\bibinfo{author}{\bibfnamefont{M.}~\bibnamefont{Chhowalla}},
  \bibinfo{author}{\bibfnamefont{H.}~\bibnamefont{Wang}},
  \bibinfo{author}{\bibfnamefont{N.}~\bibnamefont{Sano}},
  \bibinfo{author}{\bibfnamefont{K.~B.~K.} \bibnamefont{Teo}},
  \bibinfo{author}{\bibfnamefont{S.~B.} \bibnamefont{Lee}}, \bibnamefont{and}
  \bibinfo{author}{\bibfnamefont{G.~A.~J.} \bibnamefont{Amaratunga}},
  \bibinfo{journal}{Phys. Rev. Lett.} \textbf{\bibinfo{volume}{90}},
  \bibinfo{pages}{155504} (\bibinfo{year}{2003}).

\bibitem[{\citenamefont{Bernatowicz et~al.}(1995)\citenamefont{Bernatowicz,
  Gibbons, Amari, and Lewis}}]{Bernatowicz95}
\bibinfo{author}{\bibfnamefont{T.~J.} \bibnamefont{Bernatowicz}},
  \bibinfo{author}{\bibfnamefont{P.~C.} \bibnamefont{Gibbons}},
  \bibinfo{author}{\bibfnamefont{S.}~\bibnamefont{Amari}}, \bibnamefont{and}
  \bibinfo{author}{\bibfnamefont{R.~S.} \bibnamefont{Lewis}}, in
  \emph{\bibinfo{booktitle}{Lunar and Planet. Sci. Conf. XXVI Abstracts}}
  (\bibinfo{publisher}{Lunar and Planetary Institute},
  \bibinfo{address}{Houston TX}, \bibinfo{year}{1995}).

\bibitem[{\citenamefont{Sedlmayr and Kr{\"u}ger}(1997)}]{Sedlmayr97}
\bibinfo{author}{\bibfnamefont{E.}~\bibnamefont{Sedlmayr}} \bibnamefont{and}
  \bibinfo{author}{\bibfnamefont{D.}~\bibnamefont{Kr{\"u}ger}}, in
  \emph{\bibinfo{booktitle}{Astrophysical implications of the laboratory study
  of presolar materials}}, edited by \bibinfo{editor}{\bibfnamefont{T.~J.}
  \bibnamefont{Bernatowicz}} \bibnamefont{and}
  \bibinfo{editor}{\bibfnamefont{E.}~\bibnamefont{Zinner}}
  (\bibinfo{publisher}{American Institute of Physics},
  \bibinfo{address}{Woodbury NY}, \bibinfo{year}{1997}), no.
  \bibinfo{number}{402} in \bibinfo{series}{AIP Conference Proceedings}, pp.
  \bibinfo{pages}{425--450}.

\bibitem[{\citenamefont{Kroto and McKay}(1988)}]{Kroto88}
\bibinfo{author}{\bibfnamefont{H.~W.} \bibnamefont{Kroto}} \bibnamefont{and}
  \bibinfo{author}{\bibfnamefont{K.}~\bibnamefont{McKay}},
  \bibinfo{journal}{Nature} \textbf{\bibinfo{volume}{331}},
  \bibinfo{pages}{328} (\bibinfo{year}{1988}).

\bibitem[{\citenamefont{Allamandola et~al.}(1989)\citenamefont{Allamandola,
  Tielens, and Barker}}]{Allamandola89}
\bibinfo{author}{\bibfnamefont{L.~J.} \bibnamefont{Allamandola}},
  \bibinfo{author}{\bibfnamefont{G.~G.~M.} \bibnamefont{Tielens}},
  \bibnamefont{and} \bibinfo{author}{\bibfnamefont{J.~R.}
  \bibnamefont{Barker}}, \bibinfo{journal}{Astrophysical Journal Supplement
  Series} \textbf{\bibinfo{volume}{71}}, \bibinfo{pages}{733}
  (\bibinfo{year}{1989}).

\bibitem[{\citenamefont{Bernatowicz and Cowsik}(1997)}]{Bernatowicz97c}
\bibinfo{author}{\bibfnamefont{T.~J.} \bibnamefont{Bernatowicz}}
  \bibnamefont{and} \bibinfo{author}{\bibfnamefont{R.}~\bibnamefont{Cowsik}},
  in \emph{\bibinfo{booktitle}{Astrophysical implications of the laboratory
  study of presolar materials}}, edited by
  \bibinfo{editor}{\bibfnamefont{T.~J.} \bibnamefont{Bernatowicz}}
  \bibnamefont{and} \bibinfo{editor}{\bibfnamefont{E.}~\bibnamefont{Zinner}}
  (\bibinfo{publisher}{American Institute of Physics},
  \bibinfo{address}{Woodbury NY}, \bibinfo{year}{1997}), no.
  \bibinfo{number}{402} in \bibinfo{series}{AIP Conference Proceedings}, pp.
  \bibinfo{pages}{451--474}.

\bibitem[{\citenamefont{Ghiringhelli et~al.}(2005)\citenamefont{Ghiringhelli,
  Los, Meijer, Fasolino, and Frenkel}}]{Ghiringhelli05}
\bibinfo{author}{\bibfnamefont{L.~M.} \bibnamefont{Ghiringhelli}},
  \bibinfo{author}{\bibfnamefont{J.~H.} \bibnamefont{Los}},
  \bibinfo{author}{\bibfnamefont{E.~J.} \bibnamefont{Meijer}},
  \bibinfo{author}{\bibfnamefont{A.}~\bibnamefont{Fasolino}}, \bibnamefont{and}
  \bibinfo{author}{\bibfnamefont{D.}~\bibnamefont{Frenkel}},
  \bibinfo{journal}{Phys. Rev. Lett.}
  \textbf{\bibinfo{volume}{94}}(\bibinfo{number}{14}), \bibinfo{pages}{145701}
  (\bibinfo{year}{2005}).

\bibitem[{\citenamefont{Michael et~al.}(2003)\citenamefont{Michael, Nuth, and
  Lilleleht}}]{Michael03}
\bibinfo{author}{\bibfnamefont{B.~P.} \bibnamefont{Michael}},
  \bibinfo{author}{\bibfnamefont{J.~A.} \bibnamefont{Nuth}}, \bibnamefont{and}
  \bibinfo{author}{\bibfnamefont{L.~U.} \bibnamefont{Lilleleht}},
  \bibinfo{journal}{Astrophys. J.} \textbf{\bibinfo{volume}{590}},
  \bibinfo{pages}{579} (\bibinfo{year}{2003}).

\bibitem[{\citenamefont{Amari et~al.}(1994)\citenamefont{Amari, Lewis, and
  Anders}}]{Amari94}
\bibinfo{author}{\bibfnamefont{S.}~\bibnamefont{Amari}},
  \bibinfo{author}{\bibfnamefont{R.~S.} \bibnamefont{Lewis}}, \bibnamefont{and}
  \bibinfo{author}{\bibfnamefont{E.}~\bibnamefont{Anders}},
  \bibinfo{journal}{Geochim. Cosmochim. Acta} \textbf{\bibinfo{volume}{58}},
  \bibinfo{pages}{459} (\bibinfo{year}{1994}).

\bibitem[{\citenamefont{Bandow et~al.}(2000)\citenamefont{Bandow, Kokai,
  Takahashi, Yudasaka, Qin, and Iijima}}]{Bandow00}
\bibinfo{author}{\bibfnamefont{S.}~\bibnamefont{Bandow}},
  \bibinfo{author}{\bibfnamefont{F.}~\bibnamefont{Kokai}},
  \bibinfo{author}{\bibfnamefont{K.}~\bibnamefont{Takahashi}},
  \bibinfo{author}{\bibfnamefont{M.}~\bibnamefont{Yudasaka}},
  \bibinfo{author}{\bibfnamefont{L.~C.} \bibnamefont{Qin}}, \bibnamefont{and}
  \bibinfo{author}{\bibfnamefont{S.}~\bibnamefont{Iijima}},
  \bibinfo{journal}{Chem. Phys. Lett.} \textbf{\bibinfo{volume}{321}},
  \bibinfo{pages}{514} (\bibinfo{year}{2000}).

\bibitem[{\citenamefont{Warren}(1941)}]{Warren41}
\bibinfo{author}{\bibfnamefont{B.~E.} \bibnamefont{Warren}},
  \bibinfo{journal}{Phys. Rev.}
  \textbf{\bibinfo{volume}{59}}(\bibinfo{number}{9}), \bibinfo{pages}{693}
  (\bibinfo{year}{1941}).

\bibitem[{\citenamefont{Patterson}(1939)}]{Patterson39b}
\bibinfo{author}{\bibfnamefont{A.~L.} \bibnamefont{Patterson}},
  \bibinfo{journal}{Phys. Rev.} \textbf{\bibinfo{volume}{56}},
  \bibinfo{pages}{978} (\bibinfo{year}{1939}).

\bibitem[{\citenamefont{Rees and Spink}(1950)}]{Rees50}
\bibinfo{author}{\bibfnamefont{A.~L.~G.} \bibnamefont{Rees}} \bibnamefont{and}
  \bibinfo{author}{\bibfnamefont{J.~A.} \bibnamefont{Spink}},
  \bibinfo{journal}{Acta Cryst.} \textbf{\bibinfo{volume}{3}},
  \bibinfo{pages}{316} (\bibinfo{year}{1950}).

\bibitem[{\citenamefont{Fraundorf et~al.}(1989)\citenamefont{Fraundorf,
  Fraundorf, Bernatowicz, Lewis, and Ming}}]{Fraundorf89}
\bibinfo{author}{\bibfnamefont{P.}~\bibnamefont{Fraundorf}},
  \bibinfo{author}{\bibfnamefont{G.}~\bibnamefont{Fraundorf}},
  \bibinfo{author}{\bibfnamefont{T.}~\bibnamefont{Bernatowicz}},
  \bibinfo{author}{\bibfnamefont{R.}~\bibnamefont{Lewis}}, \bibnamefont{and}
  \bibinfo{author}{\bibfnamefont{T.}~\bibnamefont{Ming}},
  \bibinfo{journal}{Ultramicroscopy} \textbf{\bibinfo{volume}{27}},
  \bibinfo{pages}{401} (\bibinfo{year}{1989}).

\bibitem[{\citenamefont{Warren}(1969/1990)}]{Warren69}
\bibinfo{author}{\bibfnamefont{B.~E.} \bibnamefont{Warren}},
  \emph{\bibinfo{title}{X-ray diffraction}}
  (\bibinfo{publisher}{Addison-Wesley/Dover}, \bibinfo{address}{New York},
  \bibinfo{year}{1969/1990}).

\bibitem[{\citenamefont{Fraundorf et~al.}(2005)\citenamefont{Fraundorf, Qin,
  Moeck, and Mandell}}]{Fraundorf05JAP}
\bibinfo{author}{\bibfnamefont{P.}~\bibnamefont{Fraundorf}},
  \bibinfo{author}{\bibfnamefont{W.}~\bibnamefont{Qin}},
  \bibinfo{author}{\bibfnamefont{P.}~\bibnamefont{Moeck}}, \bibnamefont{and}
  \bibinfo{author}{\bibfnamefont{E.}~\bibnamefont{Mandell}},
  \bibinfo{journal}{Journal of Applied Physics} \textbf{\bibinfo{volume}{98}},
  \bibinfo{pages}{114308} (\bibinfo{year}{2005}).

\bibitem[{\citenamefont{Liu and Cowley}(1994)}]{Liu94}
\bibinfo{author}{\bibfnamefont{M.}~\bibnamefont{Liu}} \bibnamefont{and}
  \bibinfo{author}{\bibfnamefont{J.~M.} \bibnamefont{Cowley}},
  \bibinfo{journal}{Ultramicroscopy} \textbf{\bibinfo{volume}{53}},
  \bibinfo{pages}{333} (\bibinfo{year}{1994}).

\end{thebibliography}






\end{document}